\documentclass[10pt,conference]{IEEEtran}

\usepackage{tabularx}
\usepackage[utf8]{inputenc}
\usepackage{multirow}
\usepackage{soul}
\usepackage{xcolor}
\usepackage{xspace}
\usepackage{url}
\usepackage{graphicx}
\usepackage{listings,multicol}
\usepackage[caption=false]{subfig}  
\usepackage[export]{adjustbox}
\usepackage{amsmath}
\usepackage{textcomp}
\usepackage{listings}
\usepackage{hyperref}
\usepackage{cleveref}
\usepackage{algorithm}
\usepackage{enumitem}
\usepackage{tcolorbox}
\usepackage{booktabs}
\usepackage{siunitx}
\usepackage{pifont}
\usepackage[noend]{algpseudocode}
\usepackage{balance}
\usepackage{cite}

\PassOptionsToPackage{svgnames}{xcolor}

\title{Optimizing Duplicate Size Thresholds in IDEs}

\author{
\IEEEauthorblockN{Konstantin Grotov,\IEEEauthorrefmark{1}\IEEEauthorrefmark{2} Sergey Titov,\IEEEauthorrefmark{2} Alexandr Suhinin,\IEEEauthorrefmark{4} Yaroslav Golubev,\IEEEauthorrefmark{2} Timofey Bryksin\IEEEauthorrefmark{2}}
\IEEEauthorblockA{
    \IEEEauthorrefmark{4}\textit{JetBrains}, \IEEEauthorrefmark{2}\textit{JetBrains Research}, \IEEEauthorrefmark{1}\textit{Constructor University}\\
\{konstantin.grotov, sergey.titov, alexandr.suhinin, yaroslav.golubev, timofey.bryksin\}@jetbrains.com
}
}

\begin{document}

\maketitle

\begin{abstract}

In this paper, we present an approach for transferring an optimal lower size threshold for clone detection from one language to another by analyzing their clone distributions. We showcase this method by transferring the threshold from regular Python scripts to Jupyter notebooks for using in two JetBrains IDEs, Datalore and DataSpell.

\end{abstract}

\section{Introduction}

While clone detection is a well-established field of research, its practical application still faces open questions, in particular, the minimal size of clones that need to be detected to filter out trivial, universal code, with researchers often selecting different thresholds for the same techniques~\cite{wang2013searching, ragkhitwetsagul2018comparison}. Several works even researched the idea of using many different thresholds at once~\cite{keivanloo2015threshold, golubev2021multi}. This problem manifests itself in IntelliJ-based IDEs~\cite{kurbatova2021intellij}, such as PyCharm~\cite{pycharm}, where exact clones~\cite{roy2007survey} are underlined in the editor. 
To do  this, the current threshold was established empirically back when the system was developed. At the same time, different languages vary in their verboseness and thus may require different thresholds. Therefore, for the younger IDEs, we decided to establish new, more fitting, thresholds. However, firstly, it is very time-consuming to carry out a separate user-studies and A/B tests for each IDE and language. Secondly, existing users already got used to the current default thresholds, and changing the IDE's behavior drastically might interfere with their workflow. Thus, rather than implement a brand new threshold, the IDE development team wanted to \textit{update} and \textit{optimize} the threshold for a new language so that the amount of detected clones is roughly the same. In this paper, we describe how we did this for two new JetBrains IDEs --- Datalore~\cite{datalore} and DataSpell~\cite{dataspell} --- via comparing regular Python code and Jupyter notebooks.

\section{Approach, Evaluation, and Future Work}

\textbf{Approach.} Our core idea is to use distributions of all the detected clones within two languages or ecosystems to find such a threshold that would detect the same percentage of clones. In this work, we consider only exact clones within a single file, since they are the ones highlighted in the IDE.

To find all clones within one file, we employed the popular suffix tree based approach that is used in the platform~\cite{liu2006detecting}.  While the suffix tree is usually applied to strings, it is possible to build it for any arbitrary sequence, therefore, we use it directly on the elements of the platform's concrete syntax tree (CST)~\cite{kurbatova2021intellij}. Having obtained these elements from parsing, we iteratively go through all possible sizes for duplicates, from 3 tokens up to the half of the file's length. As a result, we get a list of all duplicates for each given size for the analyzed file.

\begin{figure}[t]
\includegraphics[width=0.9\columnwidth]{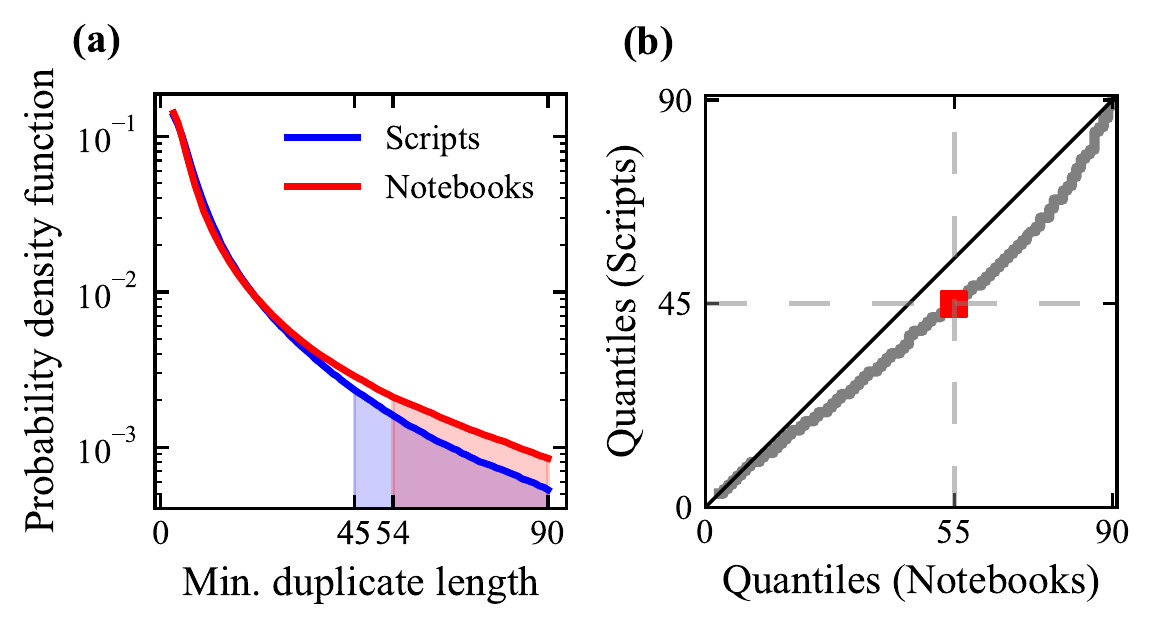}
\centering
\vspace{-0.4cm}
    \caption{(a) Probability density function of duplicate count in notebooks and scripts. (b) Quantile-Quantile plot of notebook and script clone distributions. Red square indicates the intersection of quantiles, corresponding to the same quantile's marker of different distributions.}
    \label{fig:quantiles}
    \vspace{-0.6cm}
\end{figure}

Firstly, we take the language, for which we know the optimal threshold, collect a dataset for it, and search it for clones as described above. This results in the distribution of a mean number of duplicates in a file with each tested minimal threshold. On this distribution, we can find the bin corresponding to our optimal value and calculate its quantile rank. In order to find the optimal threshold for another language, we need to repeat the same process and find the threshold that corresponds to the same quantile rank in its distribution.

\textbf{Evaluation.} We applied this approach to find the optimal value for Jupyter Notebooks for two new IDEs --- Datalore~\cite{datalore} and DataSpell~\cite{dataspell}. Research shows that code clones are frequent in notebooks~\cite{koenzen2020code} and that the code in regular Python and notebooks is different~\cite{grotov2022large}. 
We sampled 10,000 Python scripts and 10,000 Jupyter notebooks with permissive licences and 10+ stars on GitHub from our previous work~\cite{grotov2022large}. Next, we applied our algorithm, and calculated that the default PyCharm's threshold of 45 CST elements corresponds to the 95th percentile of the distribution, meaning that 5\% of all potential Python clones are underlined. The area under curve of the probability density functions (Figure~\ref{fig:quantiles}a) shows that to highlight the same percentage in notebooks, the threshold needs to be higher, at 54 elements, since Jupyter notebooks have a ``heavier tail'' of larger clones. The same can be visualized as a QQ-plot (Figure~\ref{fig:quantiles}b). You can find additional technical details and figures in our online appendix~\cite{appendix}.

\textbf{Future work.} Currently, the obtained threshold is being evaluated by the development teams of Datalore and DataSpell, we plan to carry out UX studies with users to compare the threshold's comfortability.
We believe that our pipeline, while simple, can be useful for various practical applications of clone detection. It can be used for any other family of languages, for example, for JVM-based languages.

\bibliographystyle{IEEEtran}
\balance
\bibliography{IEEEabrv,paper}

\begin{thebibliography}{10}
\providecommand{\url}[1]{#1}
\csname url@samestyle\endcsname
\providecommand{\newblock}{\relax}
\providecommand{\bibinfo}[2]{#2}
\providecommand{\BIBentrySTDinterwordspacing}{\spaceskip=0pt\relax}
\providecommand{\BIBentryALTinterwordstretchfactor}{4}
\providecommand{\BIBentryALTinterwordspacing}{\spaceskip=\fontdimen2\font plus
\BIBentryALTinterwordstretchfactor\fontdimen3\font minus
  \fontdimen4\font\relax}
\providecommand{\BIBforeignlanguage}[2]{{%
\expandafter\ifx\csname l@#1\endcsname\relax
\typeout{** WARNING: IEEEtran.bst: No hyphenation pattern has been}%
\typeout{** loaded for the language `#1'. Using the pattern for}%
\typeout{** the default language instead.}%
\else
\language=\csname l@#1\endcsname
\fi
#2}}
\providecommand{\BIBdecl}{\relax}
\BIBdecl

\bibitem{wang2013searching}
T.~Wang, M.~Harman, Y.~Jia, and J.~Krinke, ``Searching for better
  configurations: a rigorous approach to clone evaluation,'' in
  \emph{Proceedings of the 2013 9th Joint Meeting on Foundations of Software
  Engineering}, 2013, pp. 455--465.

\bibitem{ragkhitwetsagul2018comparison}
C.~Ragkhitwetsagul, J.~Krinke, and D.~Clark, ``A comparison of code similarity
  analysers,'' \emph{Empirical Software Engineering}, vol.~23, no.~4, pp.
  2464--2519, 2018.

\bibitem{keivanloo2015threshold}
I.~Keivanloo, F.~Zhang, and Y.~Zou, ``Threshold-free code clone detection for a
  large-scale heterogeneous {Java} repository,'' in \emph{2015 IEEE 22nd
  International Conference on Software Analysis, Evolution, and Reengineering
  (SANER)}.\hskip 1em plus 0.5em minus 0.4em\relax IEEE, 2015, pp. 201--210.

\bibitem{golubev2021multi}
Y.~Golubev, V.~Poletansky, N.~Povarov, and T.~Bryksin, ``Multi-threshold
  token-based code clone detection,'' in \emph{2021 IEEE International
  Conference on Software Analysis, Evolution and Reengineering (SANER)}.\hskip
  1em plus 0.5em minus 0.4em\relax IEEE, 2021, pp. 496--500.

\bibitem{kurbatova2021intellij}
Z.~Kurbatova, Y.~Golubev, V.~Kovalenko, and T.~Bryksin, ``The {IntelliJ}
  platform: a framework for building plugins and mining software data,'' in
  \emph{2021 36th IEEE/ACM International Conference on Automated Software
  Engineering Workshops (ASEW)}.\hskip 1em plus 0.5em minus 0.4em\relax IEEE,
  2021, pp. 14--17.

\bibitem{pycharm}
``{PyCharm},'' \url{https://www.jetbrains.com/pycharm/}, [Online; accessed
  16-March-2023].

\bibitem{roy2007survey}
C.~K. Roy and J.~R. Cordy, ``A survey on software clone detection research,''
  \emph{Queen’s School of computing TR}, vol. 541, no. 115, pp. 64--68, 2007.

\bibitem{datalore}
``{Datalore},'' \url{https://datalore.jetbrains.com/}, [Online; accessed
  16-March-2023].

\bibitem{dataspell}
``{DataSpell},'' \url{https://www.jetbrains.com/dataspell/}, [Online; accessed
  16-March-2023].

\bibitem{liu2006detecting}
H.~Liu, Z.~Ma, L.~Zhang, and W.~Shao, ``Detecting duplications in sequence
  diagrams based on suffix trees,'' in \emph{2006 13th Asia Pacific Software
  Engineering Conference (APSEC'06)}.\hskip 1em plus 0.5em minus 0.4em\relax
  IEEE, 2006, pp. 269--276.

\bibitem{koenzen2020code}
A.~P. Koenzen, N.~A. Ernst, and M.-A.~D. Storey, ``Code duplication and reuse
  in {Jupyter} notebooks,'' in \emph{2020 IEEE Symposium on Visual Languages
  and Human-Centric Computing (VL/HCC)}.\hskip 1em plus 0.5em minus 0.4em\relax
  IEEE, 2020, pp. 1--9.

\bibitem{grotov2022large}
K.~Grotov, S.~Titov, V.~Sotnikov, Y.~Golubev, and T.~Bryksin, ``A large-scale
  comparison of {Python} code in {Jupyter} notebooks and scripts,'' in
  \emph{Proceedings of the 19th International Conference on Mining Software
  Repositories}, 2022, pp. 353--364.

\bibitem{appendix}
``{Online appendix and replicating package},''
  \url{https://github.com/JetBrains-Research/jupyter-python-clones}, [Online;
  accessed 16-March-2023].

\end{thebibliography}

\end{document}